\documentclass[12pt]{article}
\usepackage[utf8]{inputenc}
\usepackage{amsfonts}
\usepackage{amsmath}
\usepackage{amssymb}
\usepackage{graphicx}
\usepackage{cite}
\usepackage{color}
\usepackage[left=1cm,top=1cm,bottom=1cm,right=1cm,nohead,nofoot]{geometry}

\def \be {\begin{equation}}
\def \ee {\end{equation}}
\def \bea {\begin{eqnarray}}
\def \eea {\end{eqnarray}}
\def \nn {\nonumber}

\def \rr {\raise.35ex\hbox{\small $\prime$}\kern-.17em{\mbox{\large $\imath$}}}
\def \del {\partial}
\def \dels {\partial\kern-.6em /\kern.1em}
\def \As {{A\kern-.5em / \kern.5em}}
\def \Ds {D\kern-.7em / \kern.5em}
\def \a {\alpha}

\def \b {\beta}

\def \ks {k\kern-.5em /}
\def \ls {l\kern-.5em /}

\reversemarginpar
\newcommand{\hide}[1]{}

\begin{document}
\begin{titlepage}

\begin{center}

\hfill
\vskip .2in

\textbf{\LARGE
Boundary Conditions and the Generalized Metric Formulation of the Double Sigma Model
\vskip.5cm
}

\vskip .5in
{\large
Chen-Te Ma \footnote{e-mail address: yefgst@gmail.com}\\
\vskip 3mm
}
{\sl
Department of Physics, Center for Theoretical Sciences and
Center for Advanced Study in Theoretical Sciences, 
National Taiwan University, Taipei 10617, Taiwan, R.O.C.
}\\
\vskip 3mm
\vspace{60pt}
\end{center}
\begin{abstract}
Double sigma model with the strong constraints is equivalent to the normal sigma model by imposing the self-duality relation. The gauge symmetries are the diffeomorphism and one-form gauge transformation with the strong constraints. We modify the Dirichlet and Neumann boundary conditions with the fully $O(D, D)$ description from the doubled gauge fields. We perform the one-loop $\beta$ function for the constant background fields to find low energy effective theory without using the strong constraints. The low energy theory can also be $O(D,D)$ invariant as the double sigma model. We use the other one way to construct different boundary conditions from the projectors. Finally, we combine the antisymmetric background field with the field strength to redefine a different $O(D, D)$ generalized metric. We use this generalized metric to construct a consistent double sigma model with the classical and quantum equivalence. We show the one-loop $\beta$ function for the constant background fields and obtain the normal sigma model after integrating out the dual coordinates.

\end{abstract}

\end{titlepage}

\section{Introduction}
\label{1}
Duality shows the nontrivial equivalence between two theories. It gives us a hope to unify all known theories. It is one of the important problems in the M-theory. For the ten dimensional theories, we have the T- and S-duality. The T-duality is an equivalence between different radius. We exchange the momentum and winding modes in closed string theory and the Dirichlet and Neumann boundary conditions in open string theory. The T-duality suffers from the T-fold problem in the closed string field theory \cite{Zwiebach:1992ie}. Even for the simple constant flux situation, we still find non-single valued fields because of non-isometry. It shows that the T-duality is not a well-defined transition function as gauge transformation or diffeomorphism. The S-duality is an equivalence between strong and weak coupling constants. It is a non-perturbative duality so we cannot use perturbation with the coupling constant parameter. Invalidity of perturbation gives rise to a trouble. As a familiar example, it is electric-magnetic duality of the abelian gauge theory. For the non-abelian gauge theories, we do not know how to define the electric-magnetic duality. Full understanding of the S-duality remains open. In the eleven dimensions, we combine the T- and S-duality to form the U-duality. The U-duality is expected to be the symmetry of the eleven dimensional supergravity. 

The method of solving the T-fold problem is to extend the local to global geometry. So far many low energy effective theories \cite{Ho:2008nn} are defined on the local geometry. The global geometry of the brane theory is the generalized Dirac–Born–Infeld (DBI) theory \cite{Jurco:2013upa}. The non-commutative geometry of this theory is governed by the generalized metric, which is an important element to combine tangent with cotangent bundle. It gives us a new perspective to construct the low energy effective theory or extend the T-fold understanding. This low energy effective theory also has a corresponding sigma model \cite{Jurco:2013upa, Lee:2010ey} from the new generalized metric. If we try to combine vector with one-form, we can have double geometry. This new geometry possibly be a good description to describe string theory \cite{Hohm:2011dv}. They doubled coordinates (normal and dual coordinates) to embed the T-duality rule in the $O(D, D)$ structure for the closed string theory \cite{Hull:2009mi, Hull:2009zb, Hohm:2010pp}. This extension gives a Courant bracket, which shows a way to solve the T-fold problem \cite{Aldazabal:2011nj}. Its extension helps us to define exotic brane. The source of exotic brane is non-geometric flux ($Q$- and $R$-flux). One example is the $5_2^2$-brane theory \cite{Kimura:2014upa}. This brane theory comes from the Neveu-Schwarz five-brane (NS5-brane) by performing two times T-duality. The double geometry suffers from constraints. The relaxing constraints \cite{Geissbuhler:2013uka} is a hard problem due to the generalized Lie derivative is not a closed algebra. The extension of $\alpha^{\prime}$ correction is constructed in \cite{Hohm:2013jaa}. Recent reviews of double geometry are in \cite{Hohm:2013bwa}. For the same understanding of the U-duality as the T-duality, we need to extend this double geometry to the exceptional field theory or exceptional generalized geometry \cite{Hohm:2013vpa}.

Double geometry of open string is proposed from \cite{Hull:2004in}. They use the similar ways with the closed string theory and suggest that the projectors should satisfy the boundary conditions. The gauge transformation and properties \cite{Hatsuda:2012uk} of a theory can be understood from the generalized geometry \cite{Gualtieri:2003dx}. The extension of the gauge transformation from the generalized geometry to double geometry of the ten dimensional supergravity is governed by the $F$-bracket \cite{Ma:2014kia}. The strong constraints (Removing the dependence of the dual coordinates.) of the $F$-bracket has exact one-form difference from the Courant bracket. The double sigma model with open string is also found from this gauge transformation. It already has classical equivalence and quantum equivalence at one-loop level \cite{Ma:2014vqm}. Quantum fluctuation of the string theory shows the low energy effective theory \cite{Zwiebach:1985uq}. One-loop $\beta$ function of the double sigma model for the closed string with the dilaton gives the consistent low energy effective action \cite{Berman:2007yf}. The conditions of the quantum conformal and Lorentz invariance are also shown in \cite{Avramis:2009xi}. The most interesting case of the one-loop quantum fluctuation is to simultaneously consider the fluctuation of the normal and dual coordinates, which gives us the correct equation of motion for the generalized metric \cite{Copland:2011wx}. It exactly shows the low energy effective action of the generalized metric formulation \cite{Hohm:2010pp}. The covariant version of the double sigma model is constructed in \cite{Lee:2013hma}.

Double geometry only shows the manifest formulation for the T-duality rule from the $O(D, D)$ description. We never discuss the manifest S-duality rule in this $O(D, D)$ formulation. It possibly be embedded in the $O(D, D)$ structure. The electric-magnetic duality in the electromagnetism is exchanging the electric and magnetic fields. It is equivalent to exchanging the field strength. The standard procedure of the electric-magnetic duality is the auxiliary field method. But it is not manifest. The manifest study is to use doubled gauge fields. We expect that exchanging gauge fields gives the manifest S-duality rule. It may be a new way to define the S-duality in a new perspective as the manifest T-duality rule. We naively doubled gauge fields in the boundary term. Then the boundary term does not break $O(D, D)$ invariance. Then we show the one-loop $\beta$ function to study the quantum fluctuation for the constant background fields. We can find a low energy effective action consistent with the $O(D, D)$ description. We also show the non-commutative relation at the semi-classical level (constant field strength). We point out that the boundary conditions on the $\sigma^1$ direction can be systematically constructed from the projectors. It shows more choices of boundary conditions for double sigma model. If we combine the antisymmetric background field and field strength to obtain a different $O(D, D)$ generalized metric, we can construct the low energy action from this generalized metric and scalar dilaton \cite{Ma:2014vqm}. We also use this generalized metric to build a new double sigma model. We can find the classical and quantum equivalence. We calculate the one-loop $\beta$ function for the constant background fields and obtain the normal sigma model when integrating out the dual coordinates. This double sigma model shows a different perspective to observe the manifest semi-classical non-commutative geometry. It should have more different theoretical viewpoints than \cite{Ma:2014kia}.

The plan of this paper is to first review the double sigma model in Sec. \ref{2}. Then we doubled gauge fields, compute the one-loop $\beta$ function, show the low energy effective action and non-commutative relation in Sec. \ref{3}. We also use projectors to realize boundary conditions on the $\sigma^1$ direction in Sec. \ref{4}. We combine the antisymmetric background field and field strength to form a different generalized metric, construct a double sigma model from this different generalized metric, and show classical and quantum equivalence in Sec. \ref{5}. Finally, we discuss and conclude in Sec. \ref{6}.


\section{Review of the Double Sigma Model}
\label{2}
We first review the double sigma model, then show classical equivalence for the double sigma model. At the end of the section, we write the gauge transformation.

\subsection{Classical Equivalence}
We start from 
\begin{equation}
S=-\int d^2 \sigma ~\frac{1}{2} \del^\a X^A {\cal H}_{AB} \del_\a X^B,
\end{equation}
where $\alpha=0, 1$ (We use the Greek indices to indicate the worldsheet coordinates.), $A=0, 1, \cdots, 2D-1$ ( We define the doubled target index from $A$ to $K$.), and 
\bea
{X^A}\equiv
 \begin{pmatrix} \,\tilde X_M\, \\[0.6ex] {X^M} \end{pmatrix}, \qquad 
  {\cal H}^{-1}~ \equiv ~ {\cal H}_{\bullet\,\bullet}  \,\ = \ \left({\cal H}^{AB}\right)^{-1} \ = \
  \begin{pmatrix}    g^{-1} & -g^{-1}B\\[0.5ex]
  Bg^{-1} & g-Bg^{-1}B\end{pmatrix}\;.
\eea
The index $M=0, 1, \cdots ,D-1$ (We define the non-doubled target index from $M$ to $Z$.). The normal coordinates are defined to be $X^M$ and dual coordinates are defined to be $\tilde{X}_M$. The metric field is $g$ and antisymmetric background field is $B$. We also define    
\bea
  {\cal H}~ \equiv ~ {\cal H}^{\bullet\,\bullet}  \,.
\eea
The name for ${\cal H}$  is generalized metric. For doubled target index, we use $\eta\equiv  \begin{pmatrix} 0& I \\ I& 0 \end{pmatrix}$ to raise and lower indices for the $O(D ,D)$ tensors. The index $\alpha$ is raised and lowered by the flat metric. The worldsheet metric is $(-, +)$ signature. If $h\equiv \begin{pmatrix} a& b \\ c& d \end{pmatrix}$ ( $a$, $b$, $c$ and $d$ are $D\times D$ matrices) is an $O(D, D)$ tensor, it satisfies $h^T \eta h = \eta$, where $T$ means the transpose of matrix.
The equation of motion for $X^A$ in the constant background is
\begin{equation}
\del^\a ({\cal H}_{AB} \del_\a X^B)=0.
\end{equation}
We need to eliminate the half degrees of freedom to show classical equivalence with the normal sigma model so we impose the self-duality relation
\begin{equation}
\del_\a X^A=\epsilon_{\alpha\beta}\eta^{AB}{\cal H}_{BC} (\del^{\beta} X^C).
\end{equation}
The matrix form is
\begin{align}
\begin{pmatrix}
\del_\a \tilde{X}\\
\del_\a X
\end{pmatrix}
&=
\begin{pmatrix}
0 & I\\
I & 0
\end{pmatrix}
\begin{pmatrix}
g^{-1} & -g^{-1} B \\
B g^{-1} & g-B g^{-1} B
\end{pmatrix}
\begin{pmatrix}
\epsilon_{\a\b} \del^\b \tilde{X}\\
\epsilon_{\a\b} \del^\b X
\end{pmatrix} \nn\\
&=
\begin{pmatrix}
B g^{-1} & g-B g^{-1} B\\
g^{-1} & -g^{-1} B 
\end{pmatrix}
\begin{pmatrix}
\epsilon_{\a\b} \del^\b \tilde{X}\\
\epsilon_{\a\b} \del^\b X
\end{pmatrix} \nn\\
&=
\begin{pmatrix}
B g^{-1} (\epsilon_{\a\b} \del^\b \tilde{X}) + (g-B g^{-1} B)(\epsilon_{\a\b} \del^\b X)\\
g^{-1}(\epsilon_{\a\b} \del^\b \tilde{X})-g^{-1} B (\epsilon_{\a\b} \del^\b X)
\end{pmatrix}.
\end{align}
We use two equations to represent the matrix form
\begin{align}
\label{mats}
\del_\a \tilde{X}&=B g^{-1} (\epsilon_{\a\b} \del^\b \tilde{X}) + (g-B g^{-1} B)(\epsilon_{\a\b} \del^\b X), \nn\\
\del_\a X &=g^{-1}(\epsilon_{\a\b} \del^\b \tilde{X})-g^{-1} B (\epsilon_{\a\b} \del^\b X).
\end{align}
We can determine $\del_\a \tilde{X}$ from ({\ref{mats}).
\begin{equation}
\label{selfduality}
\del_\a \tilde{X}= \epsilon_{\a\b}\,g\del^\b X+ B \del_\a X.
\end{equation}
The equations of motion can be rewritten as
\begin{align}
\del^\a 
\begin{pmatrix}
g^{-1}\del_\a \tilde{X}  -g^{-1} B \del_\a X \\
B g^{-1}\del_\a \tilde{X} +( g-B g^{-1} B) \del_\a X
\end{pmatrix}
=0.
\end{align}
We can obtain
\begin{align}
&\del^\a \big(B g^{-1}\del_\a \tilde{X} +( g-B g^{-1} B )\del_\a X\big)_M \nn\\
=& \del^\a \big(B g^{-1}( \epsilon_{\a\b}\,g\del^\b X+ B \del_\a X) +( g-B g^{-1}B )\del_\a X \big)_M \nn\\
=& \del^\a \big( \epsilon_{\a\b} B \del^\b X + B g^{-1} B \del_\a X +g \del_\a X -B g^{-1}B \del_\a X \big)_M\nn\\
=&\del^\a \big( \epsilon_{\a\b} B \del^\b X+g \del_\a X)_M
\end{align}
for the lower component of the equations of motion. It matches with the equation of motion for the normal sigma model by changing from $B$ to $-B$. The normal sigma model is
\bea
\frac{1}{2}\int d^2\sigma\ \bigg(\partial_{\alpha}X^M g_{MN}\partial^{\alpha}X^N-\epsilon^{\alpha\beta}\partial_{\alpha}X^MB_{MN}\partial_{\beta}X^N\bigg).
\eea
This sigma model can be extended to the non-constant background. The action is
\bea
\label{bulk}
S_{\mbox{bulk}}=\frac{1}{2}\int d^2\sigma\ \bigg(\partial_1X^A{\cal H}_{AB}\partial_1X^B-\partial_1X^A\eta_{AB}\partial_0X^B\bigg).
\eea
We use the strong constraints $\tilde{\partial}^M$=0 ($\partial_M\equiv\frac{\partial}{\partial x^M},\ 
\tilde{\partial}^M\equiv\frac{\partial}{\partial\tilde{x}_M}$ and $
{\partial_A }\equiv
 \begin{pmatrix} \,\tilde{\partial}^M \, \\[0.6ex] {\partial_M } \end{pmatrix}$.) and the self-duality relation
\bea
\label{con1}
{\cal H}^M{}_B\partial_1X^B-\eta^M{}_B\partial_0X^B=0
\eea
to guarantee the classical equivalence with the normal sigma model. If we consider the Neumann boundary condition on the $\sigma^1$ direction, we should put 
\bea
S_{\mbox{boundary}}=-\int d\sigma^0\ A_M\partial_0X^M
\eea 
to obtain the gauge invariance on the boundary. The one-loop $\beta$ function of this double sigma model for the constant background fields which should give the DBI model \cite{Ma:2014vqm}.

\subsection{Gauge Transformation}
 The gauge transformation is
\bea
\delta_{\xi} X^A&=&\xi^C\partial_C X^A+(\partial^A\xi_C-\partial_C\xi^A)X^C,
\nn\\
\delta_{\xi} {\cal H}^{AB}&=&\xi^C\partial_C{\cal H}^{AB}+(\partial^A\xi_C-\partial_C\xi^A){\cal H}^{CB}+(\partial^B\xi_C-\partial_P\xi^B){\cal H}^{AC},
\nn\\
\delta_{\xi} A_M&=&\Lambda_M+{\cal L}_{\epsilon}A_M,
\eea
where $\delta_{\xi}$ is the gauge transformation, $
{\xi^A}\equiv
 \begin{pmatrix} \,\tilde{\xi}_M\, \\[0.6ex] {\xi^M} \end{pmatrix}\equiv \begin{pmatrix} \,\Lambda_M\, \\[0.6ex] {\epsilon^M} \end{pmatrix}$, and ${\cal L}_{\epsilon}$ is the Lie derivative along the vector field $\epsilon$.
We assume that the gauge parameters do not depend on the worldsheet coordinates. Then the double sigma model is gauge invariant and the gauge algebra is closed under the $F$-bracket with $\tilde{\partial}^M$=0 \cite{Ma:2014kia}.
\bea
\lbrack\delta_1, \delta_2\rbrack=-\delta_{\lbrack\xi_1, \xi_2\rbrack_F},
\eea
where
\bea
\lbrack\xi_1,\xi_2\rbrack_F^A=\bigg(\xi_1^D\partial_D\xi_2^A-\xi_2^D\partial_D\xi_1^A\bigg)
-\frac{1}{2}\bigg(\xi_1^D\partial^A\xi_{2D}-\xi_2^D\partial^A\xi_{1D}\bigg)
-\frac{1}{2}\partial^A\bigg(\xi_{2D}Z^{D}{}_E\xi_1^E\bigg),
\nn\\
\eea
where
\bea
Z\equiv Z^{A}{}_B\equiv\begin{pmatrix} -1 & 0
 \\ 0 & 1  \end{pmatrix}.
\eea
The index of $Z$ is raised or lowered by $\eta$.

\section{Doubled Gauge Fields}
\label{3}
We doubled gauge fields on the boundary term in the double sigma model. Then we implement the self-duality relation and compute the one-loop $\beta$ function to find the low energy effective theory. We also discuss the semi-classical non-commutative geometry and the picture of the manifest S-duality from the doubled gauge fields.

\subsection{One-Loop $\beta$ Function}
When we doubled gauge fields, the boundary term becomes
\bea
-\int d\sigma^0\ A_B\partial_0X^B,
\eea
where
\bea
{A_B}\equiv
 \begin{pmatrix} \,\tilde{A}^M\, \\[0.6ex] {A_M} \end{pmatrix}.
\eea
The name for $\tilde{A}^M$ is the dual gauge field. 
The boundary conditions on the $\sigma^1$ direction are
\bea
{\cal H}_{AB}\partial_1X^B=F_{AB}\partial_0X^B, \qquad \delta X^A\eta_{AB}\partial_0 X^B=0,
\eea
where $F_{BC}\equiv \partial_BA_C-\partial_CA_B$, and the boundary condition on the $\sigma^0$ direction is
\bea
\delta X^A=0,
\eea
where $\delta$ is the variation.
 The boundary conditions are different from the normal case. We set $B=0$ and $g=I$ ($I\equiv$ identity matrix) to simplify the calculation without losing generality in the case of the constant background. We follow \cite{Abouelsaood:1986gd} to calculate the one-loop $\beta$ function. The variation of the boundary term is  
\bea
-\int d\sigma^0\ \bigg(A_B\partial_0X^B+\xi^B F_{BC}\partial_0X^C+\frac{1}{2}\big(\xi^B\xi^C\partial_B F_{CD}\partial_0X^D+\xi^B\partial_0
\xi^CF_{BC}\big)\bigg).
\nn\\
\eea
Then we show that the Green's function on the bulk is
\bea
\bigg({\cal H}_{AB}\partial_1^2-\eta_{AB}\partial_0\partial_1\bigg)G^B{}_C(\sigma, \sigma^{\prime})&=&iI_{AC}\delta^2(\sigma-\sigma^{\prime})
\eea
and on the boundary is 
\bea
{\cal H}_{AB}\partial_1G^{BC}-F_{AB}\partial_0G^{BC}=0.
\eea
The counter term on the boundary is
\bea
-\frac{1}{2}\int d\sigma^0\ \Gamma_A\partial_0X^A,
\eea
where
\bea
\Gamma_A=\lim_{\epsilon\rightarrow 0}G^{BC}(\epsilon\equiv\sigma^0-\sigma^{0\prime})\partial_BF_{CA}.
\eea
The $\beta$ function is defined by
\bea
\beta_A\equiv\epsilon\frac{\partial\Gamma_A}{\partial\epsilon}.
\eea
It is useful to change coordinates to solve the Green's function.
\bea
z=\sigma+\tau, \qquad \bar{z}=\sigma-\tau.
\eea
From the same procedure as \cite{Ma:2014vqm}, we can obtain 
\bea
\sqrt{\det\bigg({\cal H}+F\bigg)}.
\eea
from $\beta_A=0$.
This action has the $O(D, D)$ invariance as the double sigma model. For the non-constant background fields, we should obtain the same closed string theory from the bulk term and 
\bea
e^{-d}\sqrt{\det\bigg({\cal H}+F^{\prime}\bigg)}
\eea
from the boundary term. We define
\bea
e^{-d}\equiv\bigg(-\det{g}\bigg)^{\frac{1}{4}}e^{-\phi}, \qquad F^{\prime}\equiv\begin{pmatrix} 1 & B
 \\ 0 &1  \end{pmatrix}\begin{pmatrix} B_{MN}-F_{MN} & -F_M{}^N
 \\ F^M{}_N &F^{MN}  \end{pmatrix}\begin{pmatrix} 1 & 0
 \\ -B &1  \end{pmatrix},
\eea
where $d$ is called scalar dialton and $\phi$ is called dilaton.
When we exchange the normal gauge field and dual gauge field, perform the T-duality on the background fields and assume that the normal gauge and dual gauge fields only depend on the normal coordinates, we still obtain the consistent pull-back DBI action. The generalized metric implies the manifest T-duality and equivalence between the closed and open string parameters without the field strength. The leading order of the action on the flat background is the Yang-Mills term. If we assume that the dual gauge field is a constant field, we can obtain the normal Yang-Mills term. Alternatively, the normal gauge field is a constant field, we can obtain the dual Yang-Mills term. The situation exactly equals to the electric-magnetic duality. But we can find electric-magnetic duality manifestly from the doubled gauge fields. But the normal electric-magnetic duality only occurs in four dimensions, but doubled gauge fields should be in all dimensions. However, it is a way to realize the electric-magnetic duality manifestly. Finally, we calculate non-commutative relation at the semi-classical level (constant field strength). We first show
\bea
\langle X^A(z)X^B(z^{\prime})\rangle&=&-\frac{1}{2\pi}\bigg\lbrack{\cal H}^{AB}\ln\mid z-z^{\prime}\mid -{\cal H}^{AB}\ln\mid z+\bar{z}^{\prime}\mid
\nn\\
&&+\bigg(\frac{1}{{\cal H}^{-1}+\eta F\eta}{\cal H}^{-1}\frac{1}{{\cal H}^{-1}-\eta F\eta}\bigg)^{AB}\ln\mid z+\bar{z}^{\prime}\mid^2
\nn\\
&&-\bigg(\frac{1}{{\cal H}^{-1}+\eta F\eta}\eta F\eta\frac{1}{{\cal H}^{-1}-\eta F\eta}\bigg)^{AB}\ln\frac{z+\bar{z}^{\prime}}{\bar{z}+z^{\prime}}\bigg\rbrack
\eea
on the boundary ($z=-\bar{z}$ and $z^{\prime}=-\bar{z}^{\prime}$). It can be solved from
\bea
{\cal H}_{AB}(\partial_z+\partial_{\bar{z}})X^B-F_{AB}(\partial_z-\partial_{\bar{z}})X^B=0.
\eea
We restrict to the real $z$ and $z^{\prime}$ and denote them to be $\tau$ and $-\tau^{\prime}$. Then we can obtain
\bea
<X^AX^B>&=&-\frac{1}{2\pi}\bigg(\frac{1}{{\cal H}^{-1}+\eta F\eta}{\cal H}^{-1}\frac{1}{{\cal H}^{-1}-\eta F\eta}\bigg)^{AB}\ln\mid \tau-\tau^{\prime}\mid^2
\nn\\
&&-\frac{i}{2}\bigg(\frac{1}{{\cal H}^{-1}+\eta F\eta}\eta F\eta\frac{1}{{\cal H}^{-1}-\eta F\eta}\bigg)^{AB}\epsilon(\tau-\tau^{\prime}),
\eea
where $\epsilon(\tau)=1$ when $\tau>0$ and $\epsilon(\tau)=-1$ when $\tau<0$. We interpret $\tau$ as time. The non-commutative relation at the semi-classical level is
\bea
\lbrack X^A(\tau), X^B(\tau)\rbrack=T\bigg(X^A(\tau)X^B(\tau^-)-X^A(\tau)X^B(\tau^+)\bigg)=-i\bigg(\frac{1}{{\cal H}^{-1}+\eta F\eta}\eta F\eta\frac{1}{{\cal H}^{-1}-\eta F\eta}\bigg)^{AB}.
\nn\\
\eea
It means that we can find the non-commutative geometry between the normal and dual coordinates. The non-commutativity is governed by the field strength. If we do not have the dual gauge field, we only have non-commutativity on the normal coordinates. For the constant field strength and the $O(D, D)$ boundary conditions, we can embed the semi-classical non-commutative geometry in the $O(D, D)$ structure \cite{Polyakov:2015wna}.

\section{Boundary Conditions from the Projectors}
\label{4}
We use other approaches to obtain different boundary conditions. This way is also suitable for the doubled gauge fields.
We implement the boundary conditions on the $\sigma^1$ direction from the projectors.
The boundary conditions on the $\sigma^1$ direction are
\bea
\Pi_N{\cal H}^{-1}\partial_1 X=0, \qquad \Pi_D\partial_0X=0.
\eea
The first one is the Neumann-like boundary condition and the other one is the Dirichlet-like boundary condition. Then we will use projectors to project out the Dirichlet-like boundary condition on the $\sigma^1$ direction.
The projectors ($\Pi_N$ and $\Pi_D$) should satisfy
\bea
\label{pro}
\Pi_N^2=\Pi_N, \qquad \Pi_N+\Pi_D^T=1.
\eea
We can derive 
\bea
\Pi_D^2=\Pi_D
\eea 
from ({\ref{pro}).
The equations of motion on the boundary is
\bea
\Pi_N\bigg({\cal H}^{-1}\partial_1 X+\eta\partial_0X\bigg)=0.
\eea
Then we can obtain
\bea
\Pi_N\bigg({\cal H}^{-1}\partial_1X+\eta\Pi_N^T\partial_0X\bigg)=0.
\eea
If we want to get the Neumann-like boundary condition and remove the Dirichlet-like boundary condition, we need to assume
\bea
\Pi_N\eta\Pi_N^T=0.
\eea
The Neumann-like boundary condition is equivalent to projecting out the dual coordinates. Alternatively, we use $\eta$ to go to the dual frame. We equivalently exchange the normal and the dual coordinates. It implies that we have the Dirichlet-like boundary condition with respect to the dual frame. Then we either project out the dual coordinates with respect to the normal frame or project out the normal coordinates with respect to the dual frame to obtain the Neumann-like boundary condition.
If we use $\eta$ to transform $X$, we can obtain
\bea
X^{\prime}=\eta X.
\eea 
We can deduce
\bea
X^{\prime T}\bigg(\Pi_N+\Pi_D^T\bigg)X^{\prime}=X^T\eta\bigg(\Pi_N+\Pi_D^T\bigg)\eta X.
\eea
Then we assume 
\bea
\eta\Pi_N\eta=\Pi_D.
\eea
From this assumption, we can obtain
\bea
X^T\bigg(\Pi_N^T+\Pi_D\bigg)X.
\eea
We can use
\bea
\Pi_N\eta=\eta\Pi_D, \qquad \Pi_N\eta\Pi_N^T=0
\eea
to show
\bea
\Pi_D\Pi_N^T=0.
\eea
It shows 
\bea
\Pi_D\eta\Pi_D^T=0.
\eea
We assume
\bea
 \Pi_N= \begin{pmatrix} a& b \\ c& d \end{pmatrix}.
\eea
From $\Pi_D^T\eta\Pi_D=0$, we can obtain
\bea
&&\begin{pmatrix} 1-a& -b \\- c& 1-d \end{pmatrix}\begin{pmatrix} 0& 1 \\ 1& 0 \end{pmatrix}
\begin{pmatrix} 1-a^T& -c^T \\ -b^T& 1-d^T \end{pmatrix}
\nn\\
&=&\begin{pmatrix} -b& 1-a \\ 1-d& -c \end{pmatrix}\begin{pmatrix} 1-a^T& -c^T \\ -b^T& 1-d^T
\end{pmatrix}
\nn\\
&=&\begin{pmatrix} -b(1-a^T)-(1-a)b^T& bc^T+(1-a)(1-d^T) \\ (1-d)(1-a^T)+cb^T& -(1-d)c^T-c(1-d^T)
 \end{pmatrix}
=0.
\nn\\
\eea
The conditions are
\bea
b(1-a^T)=-(1-a)b^T, \qquad bc^T=-(1-a)(1-d^T), \qquad (1-d)c^T=-c(1-d^T).
\nn\\
\eea
From $\Pi_N\eta\Pi_N^T=0$, we can obtain
\bea
&&\begin{pmatrix} a& b \\ c& d \end{pmatrix}\begin{pmatrix} 0& 1 \\ 1& 0 \end{pmatrix}
\begin{pmatrix} a^T& c^T \\ b^T& d^T \end{pmatrix}
\nn\\
&=&\begin{pmatrix} b& a \\ d& c \end{pmatrix}\begin{pmatrix} a^T& c^T \\ b^T& d^T
\end{pmatrix}
\nn\\
&=&\begin{pmatrix} ba^T+ab^T& bc^T+ad^T \\ da^T+cb^T& dc^T+cd^T
 \end{pmatrix}
=0.
\eea
Then we can get conditions
\bea
ba^T=-ab^T, \qquad bc^T=-ad^T, \qquad dc^T=-cd^T.
\eea
From $\Pi_N^2=\Pi_N$, we can obtain
\bea
&&\begin{pmatrix} a& b \\ c& d \end{pmatrix}\begin{pmatrix} a& b \\ c& d \end{pmatrix}
\nn\\
&=&\begin{pmatrix} a^2+bc& ab+bd \\ ca+dc& cb+d^2 \end{pmatrix}
\nn\\
&=&\begin{pmatrix} a& b \\ c& d
 \end{pmatrix}
.
\eea
Therefore, we can find
\bea
a^2+bc=a, \qquad ab+bd=b, \qquad ca+dc=c, \qquad cb+d^2=d.
\eea
From the above conditions, we can obtain
\bea
b=-b^T, \qquad ba^T=ab, \qquad a+d^T=1, \qquad bc=a(1-a), \qquad c^T=-c, \qquad a^Tc=ca.
\nn\\
\eea
The above construction is exactly consistent with \cite{Hull:2004in}.
Then we show the Green's function on the boundary.
\bea
\Pi_N^T\bigg({\cal H}^{-1}\partial_1G-\eta F\eta\partial_0G\bigg)=0.
\eea
If we use the same way as \cite{Abouelsaood:1986gd} to obtain the Green's function for all projectors, we should meet trouble. The problem is that the projectors are not invertible. However, we can solve them case by case. For example, we can choose 
\bea
\Pi_N=\begin{pmatrix} 0& 0 \\ 0& 1 \end{pmatrix}
\eea
to find the DBI theory as \cite{Ma:2014vqm}.

\section{Generalized Metric Formulation}
\label{5}
We combine the antisymmetric background field and field strength to form a different $O(D, D)$ generalized metric. We use this generalized metric and scalar dialton to construct the low energy effective action. We use the same generalized metric to reconstruct our double sigma model. We also check the one-loop $\beta$ function in the case of the constant background field with the strong constraints. We consistently obtain the DBI action. At the end of the section, we integrate out the dual coordinates of the double sigma model with the strong constraints, then we can obtain the normal sigma model. It shows quantum equivalence between the double and normal sigma model exactly.
\subsection{The Low Energy Effective Action}
We consider low energy effective theory for the closed and open string theory. The open string part is based on the diffeomorphism and one-form gauge transformation. The effective action is described by the DBI action. The closed string theory can be constructed from the $O(D, D)$ structure, $\mathbb{Z}_2$ symmetry, gauge symmetry with the strong constraints and two derivative terms. Here we redefine our generalized metric by replacing $B_{MN}$ by $B_{MN}-F_{MN}$. Then we can avoid using the field strength to write the low energy effective action. The $\mathbb{Z}_2$ symmetry is
\bea
B_{MN}\rightarrow -B_{MN},
\qquad
\tilde{\partial}^M\rightarrow -\tilde{\partial}^M.
\eea
We can rewrite 
\bea
\tilde{\partial}^M\rightarrow -\tilde{\partial}^M
\eea
as
\bea
\partial_A\rightarrow  Z\, \partial_A \,.
\eea
The transformation of the ${\cal H}^{AB}$ under the $\mathbb{Z}_2$ transformation. We have
\be
{\cal H}^{AB}  \to  Z {\cal H}^{AB} Z\,, \qquad
{\cal H}_{AB}  \to  Z {\cal H}_{AB} Z \,.
\ee
Then the action can be constructed from the gauge symmetry (with the strong constraints) by using all possible $O(D, D)$ elements ($\partial_A$, ${\cal H}^{AB}$, ${\cal H}_{AB}$ and $d$) up to a boundary term and only considering two derivative terms. The action is
\bea
\label{acg}
S_2 &=& \int dx \ d\tilde x  \
   e^{-2d}\Big(\frac{1}{8}{\cal H}^{AB}\partial_{A}{\cal H}^{CD}
  \partial_{B}{\cal H}_{CD}-\frac{1}{2}
  \,{\cal H}^{AB}\partial_{B}{\cal H}^{CD}\partial_{D}
  {\cal H}_{AC}
\nn\\
  &&-2\partial_{A}d\partial_{B}{\cal H}^{AB}+4{\cal H}^{AB}\,\partial_{A}d
  \partial_{B}d \Big).
 \eea
The DBI action is 
\bea
S_1 &=& \int dx \ d\tilde x  \ e^{-d}\bigg(-\det( {\cal H}_{MN} )\bigg)^{\frac{1}{4}}.
\eea
Because of the boundary conditions are not modified from the strong constraints except for the generalized metric does not depend on the dual coordinates in $S_1$, we only rewrite the DBI action in terms of the generalized metric and scalar dilation.
We combine closed and open string to show the total action.
\bea
S_T&=&S_1+\alpha S_2
\nn\\
&=&\int dx \ d\tilde x  \ \bigg[e^{-d} \bigg(-\det( {\cal H}_{MN} )\bigg)^{\frac{1}{4}}
\nn\\
&&+\alpha e^{-2d}\bigg(\frac{1}{8}{\cal H}^{AB}\partial_{A}{\cal H}^{CD}
  \partial_{B}{\cal H}_{CD}-\frac{1}{2}
  \,{\cal H}^{AB}\partial_{B}{\cal H}^{CD}\partial_{D}
  {\cal H}_{AC}
\nn\\
  &&-2\partial_{A}d\partial_{B}{\cal H}^{AB}+4{\cal H}^{AB}\,\partial_{A}d
  \partial_{B}d\bigg) \Bigg],
\eea
where $\alpha$ is an arbitrary constant.
If we use the strong constraints, we obtain
\bea
\int dx\ \sqrt{-\det g}\bigg[e^{-\phi}\bigg( -\det(g+B-F)\bigg)^{\frac{1}{2}}\bigg( -\det g\bigg)^{-\frac{1}{2}}+\alpha e^{-2\phi}\bigg(R+4(\partial\phi)^2-\frac{1}{12}H^2\bigg)\bigg],
\nn\\
\eea
where $R$ is the Ricci scalar and $H=dB$ is the three form field strength. If we set $D$=10, it is the low energy effective theory of the D9-brane from the one-loop $\beta$ function \cite{Callan:1986bc}. 

\subsection{Double Sigma Model}
We replace $B_{MN}$ by $B_{MN}-F_{MN}$ to reconstruct our double sigma model. We discuss the classical equivalence and implement the self-duality relation at off-shell level. At the end of the section, we calculate the one-loop $\beta$ function to obtain the desirable DBI action for the constant background fields and integrate out the dual coordinates to get the normal sigma model.
\subsubsection{Action}
We replace $B_{MN}$ by $B_{MN}-F_{MN}$ to rewrite our double sigma model without using the boundary term. But we still have the boundary conditions to obtain the effects of the open string. We will show them in this section. Although we replace $B_{MN}$ by $B_{MN}-F_{MN}$, we still use $B$ in the generalized metric for simplicity. The action is
\bea
\label{dbf}
\frac{1}{2}\int d^2\sigma\ \bigg(\partial_1 X{\cal H}^{-1}\partial_1 X-\partial_1X\eta\partial_0X\bigg).
\eea
The boundary conditions on the $\sigma^1$ direction (The Neumann boundary condition) are
\bea
{\cal H}^M{}_A\partial_1X^A-\eta^M{}_A\partial_0X^A=0, \qquad {\cal H}_{MA}\partial_1X^A=0, \qquad \eta_{MA}\partial_0 X^A=0
\eea
and the boundary condition on the $\sigma^0$ direction (The Dirichlet boundary condition) is
\bea
\delta X^A=0.
\eea
We remind that the boundary conditions are not modified from the strong constraints except for the generalized metric does not depend on the dual coordinates.
\subsubsection{Classical Equivalence}
We use the on-shell self-duality relation and strong constraints to show classical equivalence with the normal sigma model. It implies that we can find the same equations of motion as the normal sigma model. The equations of motion of  (\ref{dbf}) on the bulk are
\bea
\partial_1\bigg({\cal H}_{MA}\partial_1X^A-\eta_{MA}\partial_0X^A\bigg)&=&\frac{1}{2}\partial_1X^A\partial_M{\cal H}_{AB}\partial_1X^B,
\nn\\
\partial_1\bigg({\cal H}^M{}_A\partial_1X^A-\eta^M{}_A\partial_0X^A\bigg)&=&\frac{1}{2}\partial_1X^A\partial^M{\cal H}_{AB}\partial_1X^B.
\eea
If we impose the strong constraints, we can obtain
\bea
\partial_1\bigg({\cal H}^M{}_A\partial_1X^A-\eta^M{}_A\partial_0X^A\bigg)=0.
\eea
The suitable self-duality relation is
\bea
{\cal H}^M{}_A\partial_1X^A-\eta^M{}_A\partial_0X^A=0.
\eea
The self-duality relation is equivalent to
\bea
\partial_1\tilde{X}_M=g_{MN}\partial_0X^N+B_{MN}\partial_1X^N.
\eea
The other one equation of motion is
\bea
&&\partial_1\bigg\lbrack\bigg(g-Bg^{-1}B\bigg)_{MN}\partial_1X^N+\bigg(Bg^{-1}\bigg)_M{}^N\partial_1\tilde{X}_N-\partial_0\tilde{X}_M\bigg\rbrack
\nn\\
&=&\frac{1}{2}\partial_1X^P\partial_M\bigg(g-Bg^{-1}B\bigg)_{PQ}\partial_1X^Q+\partial_1X^P\partial_M\bigg(Bg^{-1}\bigg)_P{}^Q\partial_1\tilde{X}_Q+\frac{1}{2}\partial_1\tilde{X}_P\partial_Mg^{PQ}\partial_1\tilde{X}_Q.
\nn\\
\eea
We can find the same equation of motion as the normal sigma model by using the self-duality relation to remove the dual coordinates.
\bea
&&\partial_1\bigg\lbrack\bigg(g-Bg^{-1}B\bigg)_{MN}\partial_1X^N+\bigg(Bg^{-1}\bigg)_M{}^N\partial_1\tilde{X}_N-\partial_0\tilde{X}_M\bigg\rbrack
\nn\\
&=&\partial_1\bigg(g_{MN}\partial_1X^N+B_{MN}\partial_0X^N\bigg)-\partial_0\bigg(g_{MN}\partial_0X^N+B_{MN}\partial_1X^N\bigg).
\eea
\bea
&&\frac{1}{2}\partial_1X^P\partial_M\bigg(g-Bg^{-1}B\bigg)_{PQ}\partial_1X^Q+\partial_1X^P\partial_M\bigg(Bg^{-1}\bigg)_P{}^Q\partial_1\tilde{X}_Q+\frac{1}{2}\partial_1\tilde{X}_P\partial_Mg^{PQ}\partial_1\tilde{X}_Q
\nn\\
&=&-\frac{1}{2}\partial_0X^P\partial_Mg_{PQ}\partial_0X^Q+\frac{1}{2}\partial_1X^P\partial_Mg_{PQ}\partial_1X^Q+\partial_1X^P\partial_MB_{PQ}\partial_0X^Q.
\eea
Let us consider the effects of the field strength on the bulk. Then the related terms of the equations of motion on the bulk are
\bea
&&\partial_1B_{MN}\partial_0X^N-\partial_0B_{MN}\partial_1X^N-\partial_1X^P\partial_MB_{PQ}\partial_0X^Q
\nn\\
&=&\partial_1X^P\partial_PB_{MQ}\partial_0X^Q-\partial_1X^P\partial_QB_{MP}\partial_0X^Q-\partial_1X^P\partial_MB_{PQ}\partial_0X^Q
\nn\\
&=&\partial_1X^P\partial_PB_{MQ}\partial_0X^Q+\partial_1X^P\partial_QB_{PM}\partial_0X^Q+\partial_1X^P\partial_MB_{QP}\partial_0X^Q
\nn\\
&=&\partial_1X^PH_{PMQ}\partial_0X^Q.
\eea
It implies that the field strength does not have degrees of freedom on the bulk at classical level.
On the boundary, we can have 
\bea
g_{MN}\partial_1X^N+B_{MN}\partial_0X^N=0.
\eea
It is also the normal Neumann boundary condition. We show that this double sigma model with the on-shell self-duality relation gives a consistent result with the normal sigma model. 
\subsubsection{Self-Duality Relation at Off-Shell Level}
We implement the self-duality relation at off-shell level in this section. The equations of motion on the bulk are
\bea
&&\partial_1\bigg(g^{-1}\partial_1\tilde{X}-g^{-1}B\partial_1X-\partial_0X\bigg)^M=0,
\nn\\
&&\partial_1\bigg(Bg^{-1}\partial_1\tilde{X}+\big(g-Bg^{-1}B\big)\partial_1X-\partial_0\tilde{X}\bigg)_M
\nn\\
&=&\frac{1}{2}\partial_1X\partial_M\bigg(g-Bg^{-1}B\bigg)\partial_1X+\partial_1X\partial_M\bigg(Bg^{-1}\bigg)\partial_1\tilde{X}+\frac{1}{2}\partial_1\tilde{X}\partial_Mg^{-1}\partial_1\tilde{X}.
\eea
To obtain the self-duality relation and same equations of motion as the normal sigma model, we shift $X^M$ ($X^M\rightarrow X^M+f^M(\sigma^0)$) and redefine $g$ and $B$. Then we can obtain
\bea
&&\partial_1\tilde{X}_M=B_{MN}\partial_1X^N+g_{MN}\partial_0X^N,
\nn\\
&&\partial_1\bigg(g_{MN}\partial_1X^N+B_{MN}\partial_0X^N\bigg)-\partial_0\bigg(g_{MN}\partial_0X^N+B_{MN}\partial_1X^N\bigg)
\nn\\
&=&-\frac{1}{2}\partial_0X^P\partial_Mg_{PQ}\partial_0X^Q+\frac{1}{2}\partial_1X^P\partial_Mg_{PQ}\partial_1X^Q+\partial_1X^P\partial_MB_{PQ}\partial_0X^Q.
\eea
On the boundary, the equations of motion are
\bea
\partial_1\tilde{X}-B\partial_1X-g\partial_0X&=&0,
\nn\\
Bg^{-1}\partial_1\tilde{X}+\big(g-Bg^{-1}B\big)\partial_1X&=&0.
\eea
Then we can obtain
\bea
g\partial_1X+B\partial_0X=0.
\eea
From the above discussion, the self-duality relation can be implemented at the off-shell level.
\subsubsection{One-Loop $\beta$ Function for the Constant Background Fields}
We compute the one-loop $\beta$ function for the constant background fields in this section. Finally, we will obtain the consistent DBI action. We first expand $X$ ($X\rightarrow\xi$) for the action of the double sigma model. Then we can obtain
\bea
&&\frac{1}{2}\partial_1\xi^M\big(g-Bg^{-1}B\big)_{MN}\partial_1\xi^N+\partial_1\xi^M\big(Bg^{-1}\big)_M{}^N\partial_1\tilde{\xi}_N
+\frac{1}{2}\partial_1\tilde{\xi}_M\big(g^{-1}\big)^{MN}\partial_1\tilde{\xi}_N
\nn\\
&&-\frac{1}{2}\partial_1\xi^M\partial_0\tilde{\xi}_M-\frac{1}{2}\partial_1\tilde{\xi}_M\partial_0\xi^M
\nn\\
&&+\partial_1\xi^M\xi^P\partial_P\big(g-Bg^{-1}B\big)_{MN}\partial_1X^N
+\partial_1\xi^M\xi^P\partial_P\big(Bg^{-1}\big)_M{}^N\partial_1\tilde{X}_N
-\partial_1\tilde{\xi}_M\xi^P\partial_P\big(g^{-1}B\big)^M{}_N\partial_1X^N
\nn\\
&&+\frac{1}{4}\partial_1X^P\xi^M\xi^N\partial_M\partial_N\big(g-Bg^{-1}B\big)_{PQ}\partial_1X^Q+\frac{1}{2}\partial_1X^P\xi^M\xi^N\partial_M\partial_N\big(Bg^{-1}\big)_P{}^Q\partial_1\tilde{X}_Q,
\eea
where $X^M$ and $\tilde{X}_M$ satisfy the equations of motion. The linear order of $\xi^M$ and $\tilde{\xi}_M$ disappear due to the equations of motion. We also use the strong constraints.
Because $\tilde{\xi}$ is at quadratic order, we can integrate out $\tilde{\xi}$. This integration is equivalent to the integration of
\bea
\int d^2\sigma\ \frac{1}{2}\partial_1\phi A\partial_1\phi+\phi\partial_1J,
\eea 
where $A$ ($A=A^T$) and $J$ are not related to $\phi$. Then we integrate out $\phi$, we obtain
\bea
-\int d^2\sigma\ \frac{1}{2}J\partial_1\bigg(\partial_1\big(A\partial_1\big)\bigg)^{-1}\partial_1J=-\int d^2\sigma\ \frac{1}{2}JA^{-1}J.
\eea
It is equivalent to using 
\bea
A\partial_1\phi=J.
\eea
We also use 
\bea
(\partial_1)^T=-\partial_1
\eea
and $\partial_1^{-1}$ vanishes on the boundary.
In our case, it is 
\bea
\partial_1\xi^M\big(Bg^{-1}\big)_M{}^N
+\partial_1\tilde{\xi}_M\big(g^{-1}\big)^{MN}
-\partial_0\xi^N-\xi^P\partial_P\big(g^{-1}B\big)^N{}_M\partial_1X^M=0
\eea
in the action. When we integrate by parts during the Gaussian integration process, the boundary terms will vanish due to the boundary conditions.  
Then we separate two parts to discuss. The first part is not related to $\tilde{\xi}$. We calculate all terms related to $\tilde{\xi}$ in the second part. We start from the first part.
\bea
&&\frac{1}{2}\partial_1\xi^M\bigg(g-Bg^{-1}B\bigg)_{MN}\partial_1\xi^N+\partial_1\xi^M\xi^P\partial_P\bigg(-Bg^{-1}B\bigg)_{MN}\partial_1X^N
\nn\\
&&+\partial_1\xi^M\xi^P\partial_PB_{MN}\partial_0X^N+\partial_1\xi^M\xi^P\bigg(\partial_P\big(Bg^{-1}\big)B\bigg)_{MN}\partial_1X^N
\nn\\
&&
+\frac{1}{4}\partial_1X^P\xi^M\xi^N\partial_M\partial_N\bigg(-Bg^{-1}B\bigg)_{PQ}\partial_1X^Q+\frac{1}{2}\partial_1X^P\xi^M\xi^N\bigg(\partial_M\partial_N\big(Bg^{-1}\big)B\bigg)_{PQ}\partial_1X^Q
\nn\\
&&
+\frac{1}{2}\partial_1X^P\xi^M\xi^N\partial_M\partial_NB_{PQ}\partial_0X^Q
\nn\\
&=&\frac{1}{2}\partial_1\xi^M\bigg(g-Bg^{-1}B\bigg)_{MN}\partial_1\xi^N-\partial_1\xi^M\xi^P\bigg(Bg^{-1}\partial_PB\bigg)_{MN}\partial_1X^N+\partial_1\xi^M\xi^P\partial_PB_{MN}\partial_0X^N
\nn\\
&&-\frac{1}{2}\partial_1X^P\xi^M\xi^N\bigg(\partial_MBg^{-1}\partial_NB\bigg)_{PQ}\partial_1X^Q+\frac{1}{2}\partial_1X^P\xi^M\xi^N\partial_M\partial_NB_{PQ}\partial_0X^Q.
\eea
Then we discuss the second part.
\bea
&&\frac{1}{2}\partial_1\xi^M\bigg(Bg^{-1}\bigg)_M{}^N\partial_1\tilde{\xi}_N-\frac{1}{2}\partial_1\tilde{\xi}_M\partial_0\xi^M-\frac{1}{2}\partial_1\tilde{\xi}_M\xi^P\partial_P\bigg(g^{-1}B\bigg)^M{}_N\partial_1X^N
\nn\\
&=&\frac{1}{2}\partial_1\xi^M\xi^P\bigg(Bg^{-1}\partial_PB\bigg)_{MQ}\partial_1X^Q+\frac{1}{2}\partial_1\xi^MB_{MQ}\partial_0\xi^Q+\frac{1}{2}\partial_1\xi^M\bigg(Bg^{-1}B\bigg)_{MQ}\partial_1\xi^Q
\nn\\
&&-\frac{1}{2}\xi^P\partial_PB_{QM}\partial_1X^M\partial_0\xi^Q-\frac{1}{2}g_{QM}\partial_0\xi^M\partial_0\xi^Q+\frac{1}{2}\partial_1\xi^MB_{MQ}\partial_0\xi^Q
\nn\\
&&-\frac{1}{2}\bigg(\xi^P\partial_PB_{QM}\partial_1X^M\bigg)\xi^R\partial_R\bigg(g^{-1}B\bigg)^Q{}_N\partial_1X^N-\frac{1}{2}g_{QN}\partial_0\xi^N\xi^R\partial_R\bigg(g^{-1}B\bigg)^Q{}_S\partial_1X^S
\nn\\
&&+\frac{1}{2}\partial_1\xi^MB_{MQ}\xi^P\partial_P\bigg(g^{-1}B\bigg)^Q{}_N\partial_1X^N
\nn\\
&=&\partial_1\xi^MB_{MQ}\partial_0\xi^Q+\frac{1}{2}\partial_1\xi^M\bigg(Bg^{-1}B\bigg)_{MQ}\partial_1\xi^Q-\frac{1}{2}\partial_0\xi^Ng_{QN}\partial_0\xi^Q-\partial_0\xi^M\xi^P\partial_PB_{MQ}\partial_1X^Q
\nn\\
&&+\partial_1\xi^M\xi^P\bigg(Bg^{-1}\partial_PB\bigg)_{MQ}\partial_1X^Q+\frac{1}{2}\partial_1X^P\xi^M\xi^N\bigg(\partial_MBg^{-1}\partial_NB\bigg)_{PQ}\partial_1X^Q. 
\eea
We combine two parts.
\bea
&&-\frac{1}{2}\partial_0\xi^Mg_{MN}\partial_0\xi^N+\frac{1}{2}\partial_1\xi^Mg_{MN}\partial_1\xi^N+\partial_1\xi^MB_{MN}\partial_0\xi^N
\nn\\
&&\partial_1\xi^M\xi^P\partial_PB_{MN}\partial_0X^N-\partial_0\xi^M\xi^P\partial_PB_{MQ}\partial_1X^Q+\frac{1}{2}\partial_1X^P\xi^M\xi^N\partial_M\partial_NB_{PQ}\partial_0X^Q.
\nn\\
\eea
It is exactly consistent with the second order expansion of the normal sigma model. We can redefine the one-form gauge field to absorb the constant antisymmetric background field into the one-form gauge field, we can obtain
\bea
&&-\frac{1}{2}\partial_0\xi^Mg_{MN}\partial_0\xi^N+\frac{1}{2}\partial_1\xi^Mg_{MN}\partial_1\xi^N
\nn\\
&&
-\partial_1\bigg(\frac{1}{2}\xi^N\xi^P\partial_N\partial_PA_M\partial_0X^M+\xi^N\partial_NA_M\partial_0\xi^M\bigg)
+\partial_0\bigg(\frac{1}{2}\xi^N\xi^P\partial_N\partial_PA_M\partial_1X^M+\xi^N\partial_NA_M\partial_1\xi^M\bigg).
\nn\\
\eea
We can impose the boundary conditions and integrate by parts on the boundary term, then we can get
\bea
&&\int d^2\sigma\ \bigg(\frac{1}{2}\xi^Mg_{MN}\partial_0^2\xi^N-\frac{1}{2}\xi^Mg_{MN}\partial_1^2\xi^N\bigg)
\nn\\
&&+\int d\sigma^0\ \bigg(\frac{1}{2}\xi^Mg_{MN}\partial_1\xi^N+\frac{1}{2}\xi^M\partial_0\xi^NB_{MN}+\frac{1}{2}\xi^M\xi^N\partial_MB_{NP}\partial_0X^N\bigg).
\eea
The Green's function one the bulk is
\bea
g_{MN}\bigg(\partial_0^2-\partial_1^2\bigg)G^{NP}=4g_{MN}\partial_z\partial_{\bar{z}}G^{NP}=2i\delta_M{}^P\delta^2(z-z^{\prime}),
\eea
where
\bea
\delta^2(z-z^{\prime})\equiv\frac{1}{2}\delta^2(\sigma-\sigma^{\prime}).
\eea
The solution of the Green's function on the bulk is
\bea
G^{NP}=-\frac{g^{NP}}{4\pi}\ln(z-z^{\prime})-\frac{g^{NP}}{4\pi}\ln(\bar{z}-\bar{z}^{\prime}).
\eea
Then the Green's function on the boundary is
\bea
g_{MN}\partial_1G^{NP}+B_{MN}\partial_0G^{NP}=\big(g_{MN}+B_{MN}\big)\partial_zG^{NP}+\big(g_{MN}-B_{MN}\big)\partial_{\bar{z}}G^{NP}=0.
\eea
The solution is
\bea
G^{NP}&=&{\cal H}^{NP}\ln\mid z-z^{\prime}\mid+\frac{1}{2}(g+B)^{NQ}(g-B)_{QW}{\cal H}^{WP}\ln\mid z+\bar{z}^{\prime}\mid
\nn\\
&&+\frac{1}{2}(g-B)^{NQ}(g+B)_{QW}{\cal H}^{WP}\ln(\bar{z}+z^{\prime})\mid_{z=-\bar{z}, z^{\prime}=-\bar{z}^{\prime}}.
\eea
The counter term is
\bea
\frac{1}{2}\int d\sigma^0\ \Gamma_M\partial_0X^M,
\eea
where
\bea
\Gamma_M\equiv\lim_{\epsilon\rightarrow 0} G^{NP}(\epsilon\equiv\sigma^0-\sigma^{0\prime})\partial_NB_{PM}.
\eea
Therefore, we can obtain the $\beta$ function.
\bea
\beta_M&=&{\cal H}^{NP}\ln(z-z^{\prime})+\frac{1}{2}(g+B)^{NQ}(g-B)_{QW}{\cal H}^{WP}\ln\mid z+z^{\prime}\mid
\nn\\
&&+\frac{1}{2}(g-B)^{NQ}(g+B)_{QW}{\cal H}^{WP}\ln(\bar{z}+z^{\prime})\mid_{z=-\bar{z}, z^{\prime}=-\bar{z}^{\prime}}
\nn\\
&=&2\bigg(\big({\cal H}_{YZ}\big)^{-1}\bigg)^{NP}\partial_NB_{PM}
\eea
Multiplying $\bigg(\big({\cal H}_{YZ}\big)^{-1}\bigg)$ at both sides, we obtain
\bea
\bigg(\big({\cal H}_{YZ}\big)^{-1}\bigg)^{MN}\beta_N&=&2\bigg\{\partial^P\bigg[\bigg(\big({\cal H}_{YZ}\big)^{-1}\bigg)^{MN}B_{NP}\bigg]
\nn\\
&&-\bigg(\big({\cal H}_{YZ}\big)^{-1}\bigg)^{MX}{\cal H}_X{}^W\partial_TB_{WQ}\bigg(\big({\cal H}_{YZ}\big)^{-1}\bigg)^{QP}{\cal H}_P{}^T\bigg\}.
\nn\\
\eea
The equation of motion of the DBI model is equivalent to
\bea
\sqrt{\det(g+B)}\bigg(\big({\cal H}_{YZ}\big)^{-1}\bigg)^{MN}\beta_N=0.
\eea
Although we do not show the non-constant background case, it should be consistent with the normal sigma model. We can follow \cite{Copland:2011wx} to obtain the massless closed string theory from the bulk.
\subsection{Quantum Equivalence with the Strong Constraints}
We show that this double sigma model with the strong constraints can be quantum equivalence with the normal sigma model. We integrate out the dual coordinates, then we can obtain the same normal sigma model. When we do Gaussian integration, it is equivalent to using
\bea
\partial_1\tilde{X}_P=g_{PN}\partial_0X^N+B_{PN}\partial_1X^N.
\eea
Then we show the calculation.
\bea
&&
\frac{1}{2}\partial_1X^M\bigg(g-Bg^{-1}B\bigg)_{MN}\partial_1X^N
+\partial_1X^M\bigg(Bg^{-1}\bigg)_M{}^N\partial_1\tilde{X}_N
+\frac{1}{2}\partial_1\tilde{X}_M\bigg(g^{-1}\bigg)^{MN}\partial_1\tilde{X}^N
\nn\\
&&-\partial_1\tilde{X}_M\partial_0X^M
\nn\\
&=&\frac{1}{2}\partial_1X^M\bigg(g-Bg^{-1}B\bigg)_{MN}\partial_1X^N
+\frac{1}{2}\partial_1X^M\bigg(Bg^{-1}\bigg)_M{}^N\partial_1\tilde{X}_N
-\frac{1}{2}\partial_1\tilde{X}_M\partial_0X^M
\nn\\
&=&
\frac{1}{2}\partial_1X^M\bigg(g-Bg^{-1}B\bigg)_{MN}\partial_1X^N
+\frac{1}{2}\partial_1X^MB_{MN}\partial_0X^N+\frac{1}{2}\partial_1X^M\bigg(Bg^{-1}B\bigg)_{MN}\partial_1X^N
\nn\\
&&
-\frac{1}{2}\partial_0X^Mg_{MN}\partial_0X^N
+\frac{1}{2}\partial_1X^MB_{MN}\partial_0X^N
\nn\\
&=&-\frac{1}{2}\partial_0X^Mg_{MN}\partial_0X^N+\frac{1}{2}\partial_1X^Mg_{MN}\partial_1X^N+\partial_1X^MB_{MN}\partial_0X^N.
\eea
We integrate out the dual coordinates to obtain the quantum equivalence with the normal sigma model. We can alternatively integrate out the normal coordinates with $\partial_M=0$. Then we can obtain the dual sigma model (Replacing the normal coordinates by the dual coordinates in the normal sigma model.). We can find the same situation in the generalized metric formulation at low energy level. This result shows that if we use strong constraints, the role of the dual coordinates is like an auxiliary field. The double sigma model only gives us new understanding about the duality. Without considering the duality, double geometry with the strong constraints only contains the same information as the normal sigma model. However, double geometry lets us to redefine the T-duality rule by enlarging from the $O(d, d)$ to the $O(D, D)$ structure. It possibly gives us some connection about the non-commutative geometry and the manifest T-duality.

\section{Discussion and Conclusion}
\label{6}
We discuss boundary conditions and formulate the new double sigma model by combining the field strength and antisymmetric background field. Before this paper, we only seriously consider closed string in the double geometry. We first show the full discussion of the double geometry with boundary conditions. The first discussion of the boundary conditions is the doubled gauge fields. We doubled gauge fields on the boundary. Then this theory is fully $O(D, D)$ invariant. The $O(D, D)$ invariance was broken down due to the boundary conditions. After that, we show the DBI-like action from the one-loop $\beta$ function. The difference between the DBI-like and DBI theory are the gauge fields are doubled in the DBI-like case, but not in the DBI case. If we want to obtain the DBI action, we can let the dual gauge field be a constant. The generalized metric also appears in the action. The generalized metric can govern the manifest $T$-duality rule and equivalence between the closed and open string parameters. The doubled gauge fields possibly be helpful in the manifest S-duality. On the flat background, the electric-magnetic duality is equivalent to exchanging electric and magnetic fields. The situation is the same as the doubled gauge fields. At the end of the doubled gauge fields, we show the non-commutative relation at the semi-classical level. We also use the projectors to realize different boundary conditions on the $\sigma^1$ direction. In this part, we only show conditions for the projectors. Because the projectors are not invertible, it causes a problem in considering generic cases for the one-loop $\beta$ function. However, we can choose a particular projector to go back to the DBI action. The calculation is the same as \cite{Ma:2014vqm}. Then we extend our understanding for the normal boundary conditions. We combine the field strength and antisymmetric background field to construct a double sigma model. We show the classical equivalence and implement the self-duality relation at the off-shell level. At the end of the double sigma model, we check the one-loop $\beta$ function for the constant background fields and integrate out the dual coordinates to obtain the normal sigma model.

Doubled gauge fields should be an idea framework to consider the manifest S-duality. Although it is still far from solving this problem, we already show how to realize it in the case of the flat background. The electric-magnetic duality of the non-abelian group is still not understood at this stage. We want to use the picture of the doubled gauge fields to probe the electric-magnetic duality of the non-abelian group. It should teach us more about the multiple M5-brane theory. 

We construct projectors to realize boundary conditions on the $\sigma^1$ direction. We also find the consistent DBI action in a particular projector. It should be interesting to study the one-loop $\beta$ function for the general projectors. We believe that different understanding on the boundary conditions can be obtained from the low energy effective action. These theories should be beyond the normal string theory. This method is also valid for the doubled gauge fields. However, we leave this project in the future.

We construct a double sigma model from the antisymmetric background field and field strength. The main difference are the boundary term and self-duality relation. This double sigma model do not have boundary term, but it has the consistent boundary conditions and equations of motion with the self-duality relation and strong constraints. This consistency comes from the modification of the self-duality relation. The field strength goes into the self-duality relation. It explains why we do not have the boundary term. The old double sigma model does not need the self-duality relation on the boundary, but this new double sigma model needs. We can say that the old double sigma model is a simplified version of this new double sigma model. This double sigma model naively shows that the bulk term has the effects of the one-form gauge field. But we can show that the effects of the one-form gauge field only appear in the boundary term at classical and quantum level with the strong constraints. The calculation of the one-loop $\beta$ function should be harder than the normal string theory. Furthermore, we show that this double sigma model is calculable and we also get the consistent answers. The elements of the double sigma model are the full $O(D, D)$ elements, but it does not have the full $O(D, D)$ invariance because the boundary conditions break the $O(D, D)$ invariance. If we want to have the $O(D, D)$ boundary conditions, we need to do similar construction with the doubled gauge fields. 

We can show quantum equivalence by integrating out the dual coordinates. It shows that this double sigma model with the strong constraints should be exactly equivalent to the normal sigma model beyond the one-loop level. We can also obtain the dual sigma model by integrating out the normal coordinates with $\partial_M=0$. We can use this way to observe the manifest invariance by exchanging the normal and dual coordinates as the generalized metric formulation at low energy level. Then we point out some future directions. It should be interesting to study the double sigma model without the strong constraints for quantization and one-loop $\beta$ function. Quantization should show the non-commutative relation between the normal and dual coordinates. It should be interesting to compare open string in the constant background with the closed string in the generic background. The most interesting direction of one-loop $\beta$ function should be the the low energy effective action of the double geometry. Then we can find what kind of low energy theory arisen from the fluctuation of the normal and dual coordinates. For the closed string, it is already done in \cite{Copland:2011wx}. The unsolved problem is the boundary part. From the generalized metric formulation of the closed string theory without the strong constraints, we should expect that the field strength should have effects on the bulk. Even if we consider constant background fields, it is still non-trivial because we need to consider bulk and boundary terms simultaneously. We do not have any evidences to show that this low energy effective action can be found when considering the fluctuation of the normal and dual coordinates simultaneously. But we remind that the boundary conditions are not modified from the strong constraints. It may imply that the DBI term does not have modification when considering the fluctuation of the dual coordinates. However, it should be interesting to give a new perspective for the generalized metric formulation \cite{Hohm:2010pp}. Finally, we also comment that the generalized metric, which is the combination of the field strength and antisymmetric background field can govern the semi-classical non-commutative geometry. Then we naively argue that the non-commutative geometry of closed and open theory {\it cannot} be decoupled in the double geometry. We should consider them simultaneously. This structure is not known before because we do not have the non-commutative structure on the closed string theory without doubling coordinates. However, this double sigma model should be more clearer on this point. It might be a clue that the T-duality should be more suitable on the non-commutative space. If we expect that the duality is a way to unify our theories, we should define string theory on the non-commutative space.

\section*{Acknowledgement}

The author would like to thank Dah-Wei Chiou for the discussion of the doubled gauge fields, Jun-Kai Ho for his initial collaboration and Xing Huang for the discussion of the details about the boundary conditions of the Gaussian integration. This work is supported in part by the CASTS (grant \#103R891003), Taiwan, R.O.C..

\end{document}